# A Framework to Explore the Knowledge Structure of Multidisciplinary Research Fields

**Shahadat Uddin[1], Arif Khan[1]*, Louise A. Baur[2]**

**1** Complex Systems Research Group, Project Management Program, University of Sydney, Sydney, New South Wales, Australia, **2** Discipline of Paediatrics & Child Health, and Sydney School of Public Health, University of Sydney, Sydney, New South Wales, Australia

* arif.khan@sydney.edu.au

## Abstract

Understanding emerging areas of a multidisciplinary research field is crucial for researchers, policymakers and other stakeholders. For them a knowledge structure based on longitudinal bibliographic data can be an effective instrument. But with the vast amount of available online information it is often hard to understand the knowledge structure for data. In this paper, we present a novel approach for retrieving online bibliographic data and propose a framework for exploring knowledge structure. We also present several longitudinal analyses to interpret and visualize the last 20 years of published obesity research data.

**Data Availability Statement:** All relevant data are within the paper and its Supporting Information files.

**Funding:** The authors have no support or funding to report.

**Competing Interests:** The authors have declared that no competing interests exist.

## Introduction

Scientific articles are valuable sources of information for scholarly research. Apart from the main text, their metadata hold a significant amount of information. For example, keywords and title can reveal the subject matter of the article, while citation count, impact factor etc. are used to roughly estimate significance of the article. The learnings and inferences from metadata can be treated as simple information retrieval which performs quite well when dealing with a limited number of specific articles. However, when someone wants to synthesize broader understandings on an area, such an approach has some shortcomings. One problem is that, the high volume of metadata from a large number of articles often makes it hard for readers to assimilate—resulting in an information overload [1]. Although traditional search engines and various web services (e.g., Google Scholar [2], PubMed [3], Scopus [4] and Web of Science [5]) offer search and limited analysis functions in this regard, they cannot be used effectively to analyze more complex relationships of metadata entities such as keywords, affiliation etc. So, in order to gain insights about the research field from a broader perspective, we need further analysis of metadata information to synthesize higher meaning. This can be termed as a knowledge system which abstracts, organizes and represents metadata information to general audiences. An effective knowledge system [6], therefore, acts as a valuable tool for retrieving relevant, important and summarized information [7] that are essential for different communities of stakeholders in today's knowledge-driven society. For example, this can bring a competitive advantage to stakeholders by choosing suitable field for investment, help aspiring researchers to choose a promising topic and guide funding agencies as well as policy makers in distributing funds. In this paper, we present a novel framework in order to obtain a summary picture of longitudinal research trend from web mining of bibliographical data.

Considerable amount of research has been done in organizing knowledge systems from bibliographic data. For example, metadata that contain evaluative entities [8] (e.g., paper, author, journal and affiliation) are utilized to measure scholarly impacts [9] such as identifying influential players in a specific scientific field using author citation analysis [10, 11] or ranking prestigious journals using journal citation networks [12]. Also metadata containing knowledge entities such as keyword, topics, subject categories, key method etc. act as a carrier of knowledge units in scientific articles [9] and facilitate knowledge discovery [8]. Several methods can be found which deal with the analysis of bibliographic data containing time-stamp markers (i.e., longitudinal data such as publication year, funding duration etc.). These methods [13–17] are mostly being focused on the particular contexts of trend analysis or link prediction and often employ generic statistical methods such as frequency analysis and regression analysis in various contexts to understand the research intensity and forecast future trends. Such approaches are often applied in citation analysis to reveal various scholarly impacts such as prominent articles, authors' expertise and future collaborations [18]. Some higher level approaches that consider the relation between metadata including co-word analysis [19] of entities (keywords, citation etc.), knowledge maps [20] and gene-word analysis [21] are often being used to trace the relationships between the evolution of research fields [22, 23]. The application of social network theories to bibliographic data might be considered as a more recent phenomenon which takes the complex relationships between entities into account [19]. For example, scholars use network theories on keywords to analyze the knowledge structure of research domains by defining keywords as actors and co-occurrences between keywords as ties between them. These methods can reveal different aspects of knowledge structure. However, they have some limitations at presenting the complete picture of a multi-dimensional research domain such as obesity, especially when acquiring bibliographic information from large unstructured or semi-structured data and organizing them [24]. In our proposed framework, we introduce a systematic set of analysis methods that can address this issue.

In this paper, we apply this framework in the context of the obesity domain and focus on keyword structure as the principal bibliographic entity. Keyword is chosen firstly in order to keep the paper within concise limits. Secondly, keywords represent different subtopics within obesity, which in turns helps us to understand complex interrelations and trends in those subtopics. Also, we chose the domain of obesity because it is an example of a vast research field comprising many subtopics that have significantly interacted and evolved over the past 20 years. As a major health concern worldwide, obesity was formally recognized as global epidemic in 1997 [25]. The prevalence has nearly doubled between 1980 and 2008, with approximately 1.4 billion (35%) adults worldwide considered overweight and more than half a billion (11%) obese, and over 40 million preschool overweight children [26]. The cause of, and solutions to, obesity are complex, highly interconnected and embedded within a complex network of socio-economic, cultural, medical, biological and environmental factors [27–29]. Over the past few decades, countries have undergone intense globalization and become more urbanized. Local and international migration of different socio-ethnic groups, changes in food production and availability, and changes in physical activity infrastructure have all helped spread obesity to low and middle income countries, where the problem of undernutrition continues [30]. Researchers have extensively studied the internal biological factors involving the delicate coordination of energy intake and expenditure [31]. They have made tremendous progress in defining the complex pathways that can explain obesity [32]. These underlying dynamics of different obesity sub-domains have been documented and studied in research articles. The main goal of our study is to analyze the keywords of these research articles in order to explore how effectively they can explain these evolutions in understanding obesity as well as revealing any insight that might be otherwise overlooked.

## Research Methods

In this section, we first introduce the framework and discuss its different functional parts. In the later part, we implement our research framework on particular research domain (i.e., obesity) and describe specific adaptation methods of the framework in the implementation.

### Proposed framework

Our proposed framework consists of two major parts. The first part deals with retrieving unorganized or semi-organized data from online sources and then organizes them in structured fashion. This makes the data ready for further analysis for the next part. The analysis part is executed in several steps to discover different sorts of information from the data. Fig 1 shows summarized steps of the framework. Each of the parts is described in details below.

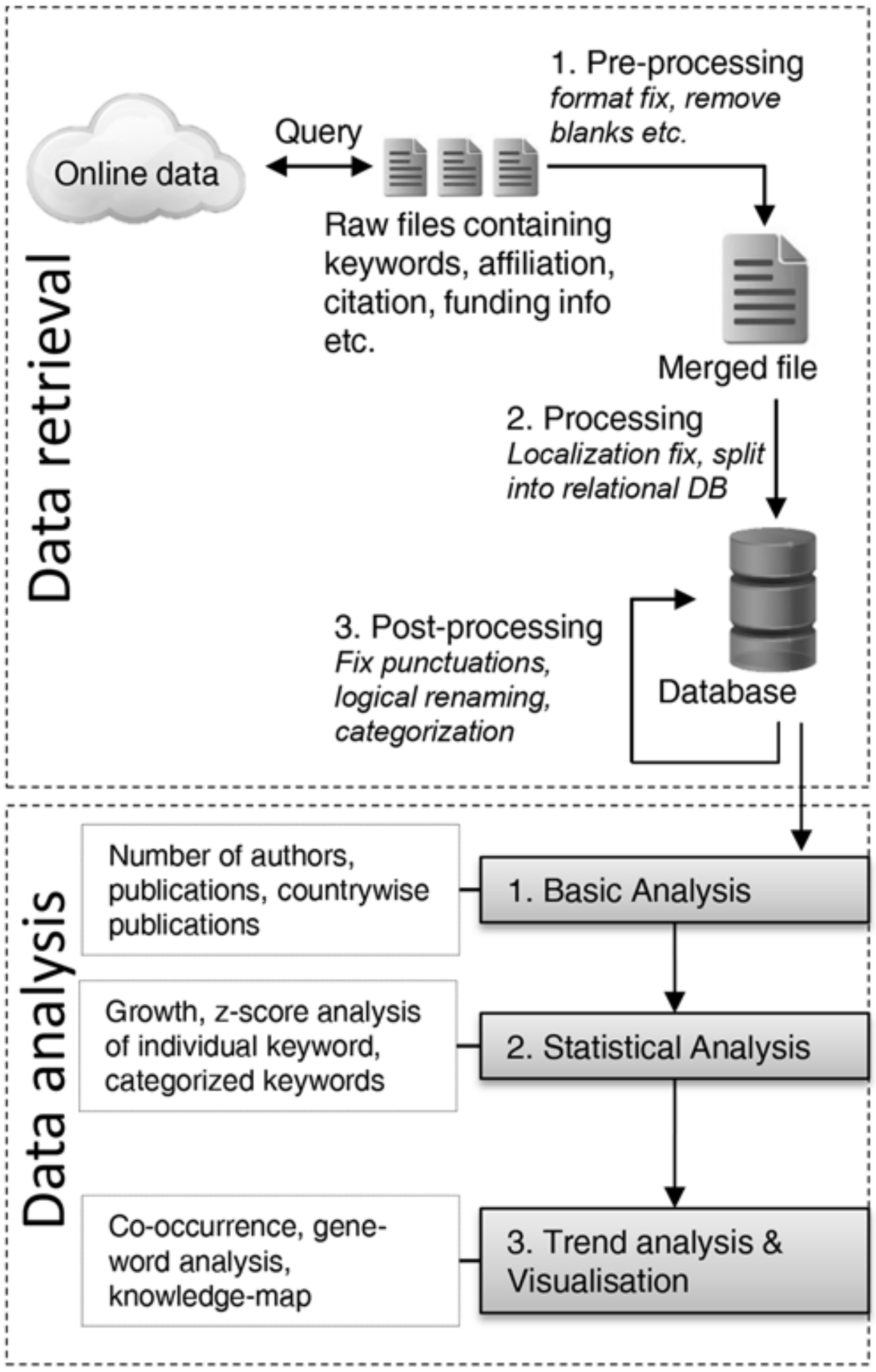

**Fig 1. Summarized research framework.**

**Data retrieval.** Different online sources, institutional libraries and web services provide bibliographic data for general users. Although these repositories contain vast amount of metadata, their web interfaces normally provide basic search and analysis function like simple frequency and trending keywords statistics etc. For most users, this information is quite sufficient but when utilized fully this vast amount of background metadata has potential to reveal the knowledge structure of the complete research field. For that, first part of the research framework concentrates on retrieving the complete metadata on particular research topic. Normally most web services let user download the metadata in tabular (i.e., excel) or structured (i.e., xml) format. However, some restrictions still apply on data usage and access policy. Many web services have subscription policy and access and downloading of metadata are limited to individually subscribed users. Many higher educational and research institutions also have group subscription and users can download the metadata within institutional network. However, this data normally do not come all at once. One important reason behind that is the security which prevents users from exhausting web resources (e.g., denial of service or DoS attack [33]). So, to get the complete metadata, our framework follows a approach which downloads the metadata sequentially either by manual querying (if download number is small) or using a script that follows the request limit guidelines setup by the web service policy. The actual method depends on the type of web service used to download the metadata.

Next, the downloaded metadata files that normally come in comma separated, XML, text or excel file format are merged together by means of a script or software in a quite straightforward way. The merged file is then inspected in several steps for formatting inconsistencies and ambiguous entries. At **pre-processing** part, the file is inspected to make sure that all records are valid. Metadata that have missing titles, empty author names or having other essential fields empty—are removed from the file. Also the file is searched for duplicate entries by matching the unique identifiers like paper titles, DOI (Digital Object Identifier) number etc. Then, in the main **processing** part, the file is inspected again to normalize entries such as renaming same words with different spelling. Due to localization of spelling of papers from different locations, same information can be present throughout the file in different formats. For example, accented letters '*á*' becomes '*a*', *Brasil* is renamed to *Brazil* etc. After this processing, files are saved into a database. The database structure is relational i.e., divided into different tables like authors, keywords, paper information, affiliation etc. that are connected by relational keys. This greatly reduces the overall size of data and makes complex analysis using database queries faster.

Finally, the last part of filtering is **post-processing** which mostly deals with the fine-tuning of entries. Although the data is organized into database, some noise is still likely to be present in forms of misspelled or synonymous entries. For example, same keyword can have several synonyms or can be spelled differently (specially chemical or compound names) using different punctuation marks. These synonymous entries are renamed to the standard name by searching into the database using script or software. As this fine-tuning method is quite exhaustive and it is quite impossible to make the dataset fully noise-free, the amount of filtering that are needed will depend on the actual dataset.

**Data analysis approaches and methods.** This part of the framework focuses on understanding knowledge structure by retrieving relevant and summarized information from data organized in previous part. The knowledge structure reveals the evolution of sub-domains within a particular research area by applying different statistical and visualization methods. In our research, we applied the methods in a hierarchical fashion starting from generic to more specific focus sequentially. These methods are divided in three broad criteria—each of which reveals different dynamics of the overall research trend; thus, providing a complete picture of overall knowledge structure.

In the **first** part of analysis, our framework focuses on revealing vital information about the overall research domain like magnitude of research involved, distribution of research across countries and the number of authors involved. To do that, we applied basic frequency based analysis methods on the dataset by executing database queries over the relational entities. In the **second** part, we applied statistical analysis method on the dataset to understand how different sub-domains of the research area have grown over the years. We focused on two statistics based methods—growth analysis and z-score. These methods are applied on individual author keywords that resemble the core topic of the paper. This in turns reveals the growth of different topics or keywords of the research area. These statistical methods are also applied similarly on grouping of the keywords into different categories that can reveal dynamics of sub-domains of research field. We have briefly described these two statistical methods and concept of categorization below.

**Growth analysis** test aims to find relative expansion or growth of particular key-topics of research fields denoted by individual keywords. It can detect sudden bursts or declines of keyword occurrence which often give an indication of major milestone, discovery or failure of a research topic. Growth analysis works by finding relative increment or decrement of frequency statistics of keywords over a certain period of time and expresses how much time, as a percentage, the statistic under consideration had increased or decreased compared to its previous value. The equation is as follows:

$$\mathrm{Growth} \quad = \quad \sum_{i=1}^{n-1} \frac{(f_{i+1} - f_i)}{f_i} \times 100\% \tag{1}$$

Where $f_i$ is frequency of a particular keyword in the $i^{th}$ time segment.

**Z-score test** is a standardized score that is used to find out the probability of occurrence of a certain topic. At first, it draws a frequency distribution graph from given a set of values (e.g., frequency of keywords) and calculate the mean and standard deviation. Then z-score utilizes these basic parameters to translate the actual frequency distribution to a standard normal distribution having mean and standard deviation of 0 and 1 respectively. The mathematical equation is:

$$z = \frac{X - \bar{X}}{s} \tag{2}$$

Where, $X$ is the value to translate, $\bar{X}$ is the mean and $s$ is the standard deviation. The resultant z-score provides a common framework to compare different set of values. Mathematically, z-score indicates the probability of occurrence. The greater the z-score the less the probability of occurrence. In our framework, a keyword having a high z-score (in terms of frequency) indicates that there is less probability of the occurrence of that keyword.

**Categorization** provides a way to group keywords that in turns represent a sub-domain within a larger research area. Individual keyword entity can reveal dynamics of a particular research topic. But it is often hard to understand the evolution of broader sub-domains of the research and how they shape the research over time. Categorizing keywords of same sub-domain and the application of above-mentioned growth and z-score methods can potentially reveal these dynamics. For our framework, keywords are categorized using a classification scheme which is usually represented as hierarchical or tree structure where top level elements represent broader category and lower level elements represent more specific category or sub-topics. To categorize a keyword, the categorization tree is searched for that particular entry. When the keyword is found, its upper level is considered as the category. It should be noted that usually there are multiple upper levels present in the categorization scheme—which represent specific to more generic domain or category as one goes to upper levels. How generic the categorization will be actually depends on the volume of research domain. Also, there are different classification scheme available for different research domain, for example, ACM Computing Classification System [34] for computing field, MeSH classification system [35] for medical science, Mathematics Subject Classification (MSC) [36] etc. Choice of exact classification system also depends on the actual research context. We specifically discussed the classification scheme that we followed for our context later in this section.

Finally, in the **third** part of analysis in our research framework, we focused on exploring the longitudinal relation and interaction between the topics and tried to visualize them graphically. The core idea is that, the topics of any research domain are related and to understand how the research trend changes, we need to look at those interrelations between the topics. For this analysis, we looked at which keywords have appeared together in a paper (co-occurrence) and made a graph based on it where nodes represent keywords and edges represent co-occurrence of keywords. The framework then applies social network measures on these keyword graphs to understand the relation between them and how they changes over the years. Also in this part, the framework analyzes the formation of keywords from a lexical perspective called Gene-word analysis and finally introduces the concept of visualization using knowledge map. These topics are described in short below.

## Descriptive measures of keyword co-occurrence network

**Average weighted degree.** This is a measure of average connection strength between any two nodes of a network [37]. For our keyword co-occurrence network, it means the average of frequency of all keyword pairs. For example, if A-B keyword pair co-occurred 10 times, B-C co-occurred 20 times and A-C co-occurred 0 times, then the average weighted degree of the network made by A, B and C will be 15. An increasing (over time) average weighted degree value for a keyword pair indicates that, on average, more research efforts have been given on the field(s) indicated by those keywords.

**Network density.** Network density is a measure of compactness [38]. A network becomes more compact when there are more links between its member nodes. Similarly, it becomes sparse when most of the nodes do not have any link between them. Mathematically, network density quantifies the number of links between any two keyword pairs over all possible pairs. A greater graph density value indicates that more inter-disciplinary research has been carried out in this field. However, this only represents, whether or not, research is carried out on two keywords simultaneously.

**Modularity.** This indicates the strength of grouping in a network [38]. Many networks show a grouping tendency. This means that network members are divided into different communities. A high modularity means nodes of the same community have more ties between them and nodes of different communities have fewer ties in between. Modularity index ranges between 0 and 1. A higher value means that grouping tendency is higher within keywords representing there are fewer co-occurrences between keywords of different groups and more co-occurrences between keywords of the same group.

**Clustering co-efficient.** This indicates the degree to which the nodes of a graph cluster together [38]. Stemming from the social network concept "*all of my friends know each other*" or "*friend of my friend is also my friend*", the clustering co-efficient measures the extent to which this idea fits for the network. For our co-occurrence network, this measure can be re-worded as like: if two keywords A and B are studied together (i.e., they are found in the same article) then it indicates that they are related in some way. Now if keywords B and C are also studied together then there is a strong possibility that keywords A and C will be studied together. Therefore,

A, B and C will form a triangle when they are visualized in a co-occurrence network. A clustering co-efficient with this perspective indicates the number of these triplets over all possible triplets representing the probability of research on any keyword pair or triplet.

**Gene-word.** A gene-word is one of the highly used words that either serves as a root of, or helps to form, many other keywords through cross-references and hybridization (i.e., being utilized as a suffix or a prefix) [21]. By following the procedure as described by Wu el al. [21], we formulated and followed three principles to extract gene-words. First, a gene-word must have a frequency of appearance of over 100 times and be ranked amongst the top-100 keywords. Second, a gene-word must be related or cross-referenced to a minimum of 10 other core keywords, each with a frequency higher than 10 and, at an aggregate level, accounting for over 20% of all keyword frequencies. Third, a gene-word must represent a priority issue in the underlying research domain.

**Data visualization using knowledge map.** A knowledge map is a discipline-specific visual representation of knowledge [19]. By considering the relative positions and density of nodes (i.e., keywords) in a network (i.e., keyword co-occurrence network), two dimensional knowledge maps can be created by using the VOSviewer which is a computer program primarily intended to be used for mapping, analyzing and exploring different types of networks (e.g., bibliometric network) [19]. In these maps, a shorter distance between any pair of keywords provides an indication of larger number of co-occurrences of those keywords (and vice versa). The font size of a keyword depends on the number and strength of its connections with the neighborhood keywords and their level of connectivity in the network. Each point in a map has a color, somewhere in between blue and red, that depends on the density of keywords at that point. The range of this color is from blue (of score 0) to green (of score 1) to red (of score 2). The larger the number of keywords in the neighborhood of a point and the higher the connectivity of the neighboring keywords, the closer the color of the point is to red. Conversely, the smaller the number of keywords in the neighborhood of a point and the lower the connectivity of the neighboring keywords, the closer the color of the point is to blue. These maps are particularly important in analyzing the fundamental structure of knowledge that is represented by keywords. For a detailed description of technical implementation and algorithm used in developing this software, see Van Eck and Waltman [39].

## Applying framework on obesity research

In this section we illustrate how this research framework can be adapted and applied on a particular research domain i.e., obesity. We also discuss specific considerations and modifications needed to apply the proposed framework on actual context.

**Data source selection and knowledge generation.** Our framework is applied on obesity research dataset retrieved from Scopus which is an online bibliographic database by Elsevier. We chose it over other online sources (e.g., Google Scholar) because of its wide coverage of medical articles, consistent accuracy, filtering options and provision for downloading complete metadata as comma-separated file(s). Scopus contents are available via subscription and there is a limit on maximum records that can be downloaded at once. The contents were accessed and downloaded manually through the University of Sydney Library's access to Scopus. The constraints for querying obesity related articles from Scopus web interface were set in this way—(i) consider only journal articles; (ii) published in English; (iii) published between year 1993 and 2013 inclusive; and (iv) the term "*obesity*" was included in the title or in author keyword list.

This resulted into 120,382 articles that were downloaded sequentially into different comma-separated files. For each paper, the metadata that were collected include: title, author keywords, author name and affiliation, publication year, citation count, journal name, funding details and identification number. We developed custom data-mining software to automate most of the processing and filtering of the web-data as per the specification of the framework.

In the pre-processing step of filtering, we manually looked for formatting inconsistencies such as encoding error, incomplete or corrupted data in the downloaded raw files. Some entries were duplicated across multiple files. These files were then inspected by software to remove those duplicate entries. Articles missing essential information (e.g., author information, year or citation count) were also discarded at this stage. Then filtered data from multiple files were merged into a single file. The output file was then inspected to normalize the entries. Because the same information may be presented throughout the file in different spellings, we programmatically fixed these spelling inconsistencies. For example, accented letters were renamed (e.g., '*á*' became '*a*' and *Brasil* was renamed to *Brazil*). After that, the data were transferred to a relational database for efficient and complex analysis. In the post-processing step, we focused on normalizing keywords with different spellings often due to misspelling and compounding of multiple words but have the same meaning. Some keywords also have a standard name from the scientific community. For example, keywords *calcifediol*, *calcidiol*, *25-hydroxyvitamin D*, *25-hydroxy vitamin D* and *25(OH) vitamin D* all refers to the same chemical, but are lexicographically different from the software perspective. We manually checked for such synonyms in most frequent keywords. This list was given to software which automatically merged the database entries. Approximation algorithms were also used by the software to merge keywords that looked different due to leading or trailing punctuation marks (e.g., *hypertension* and *hyper-tension*). Finally, we considered articles that were published between 1993 and 2012 inclusive and for which we had information about the author-provided keywords. Articles published in 2013 were not considered as the indexing was not completed in Scopus at the time of dataset downloading. The final dataset consisted of 66,567 articles.

**Data analysis of Obesity dataset.** According to the framework, we applied analysis methods on dataset to understand the knowledge structure in obesity research. In the first part of analysis, we used *basic statistics* of entities to provide a quick overview of obesity research. We explored issues such as how the research is expanding, which countries have been at the forefront of the research, and the amount of scientific workforce that is engaged in obesity research. We looked at the publication count and authors' affiliation data to answer these questions.

In the **second part** of the analysis, we focused on exploring important obesity research topics using different statistical analyses of the author provided keywords. Each article had around 5–7 keywords and, to increase accuracy in detecting topics, we merged synonymous keywords during the data retrieval stage discussed above. In order to understand the importance or significance of keywords statistically, frequency, growth and z-score analysis were applied on individual obesity keywords according to the framework. We considered $p < 0.01$ level for z-score significance.

Furthermore, the keywords were grouped into logical categories such as "*endocrine system diseases*" and "*cardiovascular diseases*" etc. to understand the trends of obesity sub-domains which might be unnoticed in individual keyword scores. Categorization was undertaken following the MeSH (Medical Subject Heading) database [35] provided by the U.S. National Library of Medicine. The classification covers the field of medicine and related sciences. The keywords are classified logically in a hierarchical fashion from generic to more specific terminologies. For example, keyword *gastroplasty* is categorized as *Surgery-> digestive system surgery-> gastroplasty*. The complete MeSH database is available to download with this hierarchical structure as an xml file. We developed software to automatically look categorization entries for keywords and save the categorization data in the database. We considered the top two levels of classifications from the hierarchy. The results were later checked manually.

Finally, in the **third part** of analysis we looked further into the relation between the keywords or topics according the measure described in the framework. It is obvious that many of the topics of obesity research will be highly related. Obesity is highly associated with different diseases, their causes, diagnostic methods and treatments are also closely related and are often researched together. Also, two obesity related keywords can individually occur more but together they might appear less. The implication may be either that the topic represented by these keywords has not much prospect or the topic might not be studied widely and, hence, there are potential research opportunities. So it might be wise for researchers looking for trending research topics to choose those keywords whose individual and combined co-occurrence score is high. Therefore, we applied network measures and statistical methods as described above to the co-occurrence network to explore how keywords expressing the different topics have occurred together in the articles' metadata and how this relation has changed over time. This part of the analysis also involves "Gene-word" analysis that is utilized to understand the syntactical relation of medical terminologies whereby a core keyword is derived into other keywords, making the topic divided into smaller branches. For example, a core keyword *"lipid"* had derived into different related keywords e.g., *dyslipidemia*, *hyperlipidemia*, *phospholipids*, *plasma lipids* etc. Finally, according to the framework we generated the "knowledge map", a visual representation of keyword co-occurrences to help in understanding the structure of obesity topics.

## Results

For cross-time analysis and ease of illustration, we divided our research dataset into four clusters or time slots based on the publication years of articles. These clusters are: 1993–1997 (cluster 1), 1998–2002 (cluster 2), 2003–2007(cluster 3) and 2008–2012 (cluster 4). All analyses were applied on these four clusters as well as overall 20 years data.

### Basic statistics and network characteristics of the research dataset

**Number of publications.** According to our research dataset, during the last 20 years (1993–2012) 66,567 papers were published in the area of obesity research. The detailed breakdown into time periods is shown in Fig 2. It is evident that since 1993 there had been a continual increase in the number of published research papers in the obesity research. This indicates

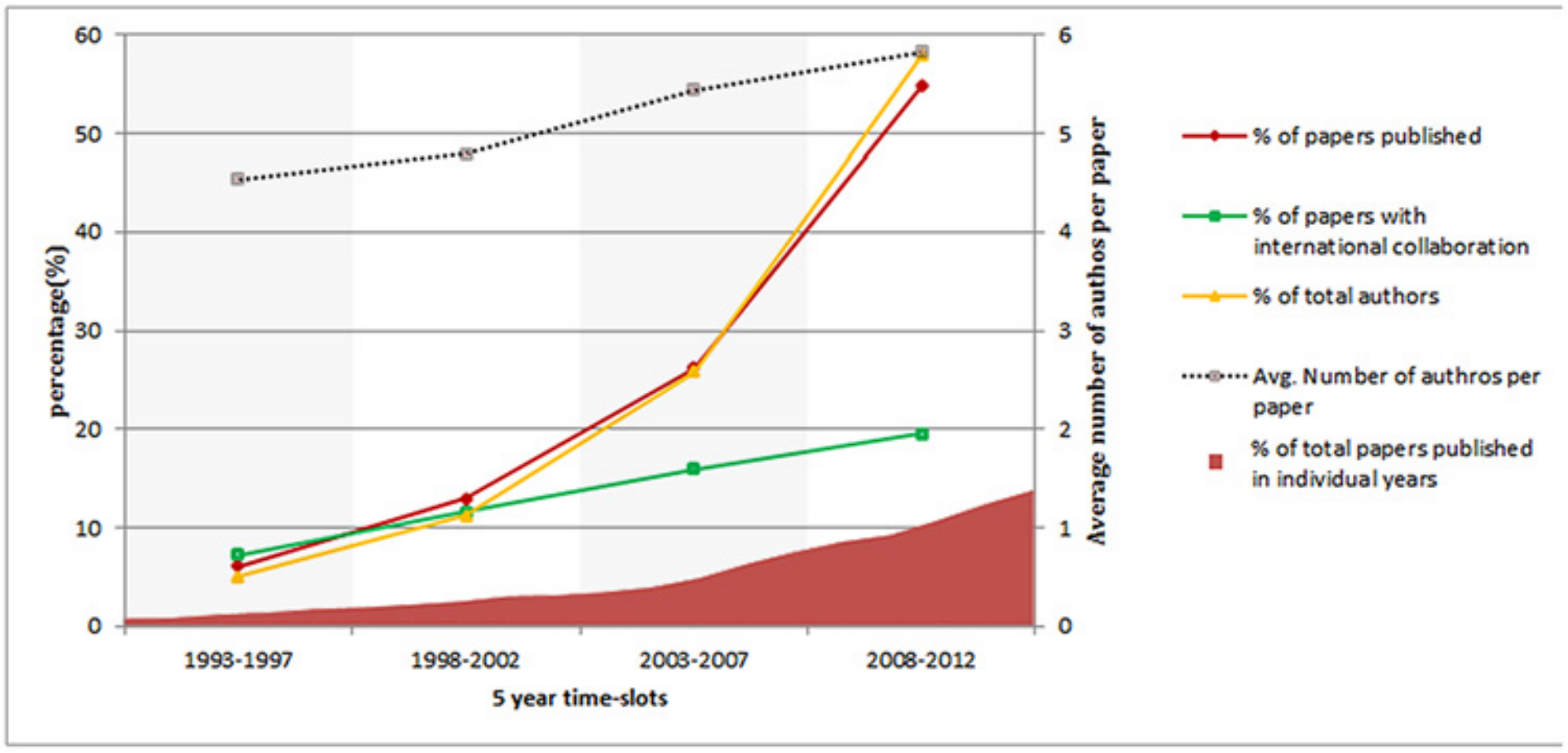

**Fig 2. Basic statistics of published papers (number of published paper, number of authors, international collaboration).**

**Table 1. Top-20 frequently occurring keywords.**

| ID | 20 years | 1993–1997 | 1998–2002 | 2003–2007 | 2008–2012 |
|---|---|---|---|---|---|
| 1 | **Body mass index** | Insulin | Body mass index | Body mass index | Body mass index |
| 2 | **Insulin resistance** | Insulin resistance | Leptin | Insulin resistance | Metabolic syndrome |
| 3 | **Diabetes** | Weight loss | Insulin resistance | Morbid obesity | Diabetes |
| 4 | Metabolic syndrome | Hypertension | Diabetes | Diabetes | Insulin resistance |
| 5 | **Children** | Body mass index | Hypertension | Metabolic syndrome | Children |
| 6 | **Hypertension** | Diabetes | Weight loss | Bariatric surgery | Adolescents |
| 7 | **Type 2 diabetes** | Type 2 diabetes | Morbid obesity | Children | Type 2 diabetes |
| 8 | **Morbid obesity** | Body composition | Insulin | Weight loss | Bariatric surgery |
| 9 | **Weight loss** | Morbid obesity | Type 2 diabetes | Hypertension | Hypertension |
| 10 | **Leptin** | Risk factors | Children | Leptin | Weight loss |
| 11 | Bariatric surgery | Blood pressure | Risk factors | Type 2 diabetes | Physical activity |
| 12 | Adolescents | Adipose tissue | Body composition | Risk factors | Risk factors |
| 13 | **Risk factors** | Leptin | Adipose tissue | Adolescents | Inflammation |
| 14 | Physical activity | Exercise | Exercise | Physical activity | Adipose tissue |
| 15 | **Adipose tissue** | Children | Bariatric surgery | Body composition | Leptin |
| 16 | Insulin | Diet | Blood pressure | Adipose tissue | Morbid obesity |
| 17 | **Body composition** | Body weight | Diet | Gastric bypass | Epidemiology |
| 18 | Epidemiology | Cholesterol | Physical activity | Insulin | Exercise |
| 19 | **Exercise** | Vertical banded gastroplasty | Coronary artery disease | Epidemiology | Adiponectin |
| 20 | Inflammation | Zucker rats | Epidemiology | Exercise | Body composition |

Bold keywords have appeared in all four time periods (could be in different numerical order) as well as in the top-20 list of the aggregated time (i.e., 20 years) that obesity, as a research domain, has been gaining more importance to researchers over the time.

**Number of authors.** There has been a steady increase in the number of authors per article over time. The statistics of the number of authors per paper in different time periods is illustrated also in Fig 2.

**Frequently used keywords.** In total, there were 327,879 keywords that appeared in 66,567 papers published over 20 years. On average, each paper had approximately 5 keywords. The top 20 most frequent author keywords in different time periods are given in Table 1. '*Body mass index*' is the most frequent keyword over the data collection period. Of the top 20 keywords for the aggregated time (i.e., 20 years), 13 keywords appeared in the top 20 list for each of the four time periods.

**International collaboration.** There were 11,185 articles out of 66,567 (16.8%) that had authors from more than one countries. The international collaboration statistics are shown also in Fig 2 along with other statistics. It reveals that there has been a sharp growth in the international collaboration of obesity research over time.

**Top-countries publishing most papers.** Authors from around 130 countries were recorded as publishing papers on the topic of obesity over the last 20 years. The publication statistics for the top 20 countries in the 4 time-slots and the global distribution of country-wise publications are shown in Fig 3. This graph shows that US had the largest numbers, followed by the UK, Japan, Italy, Canada, Australia, Germany and France.

**Network statistics of keyword co-occurrence network.** The network statistics of keyword co-occurrence networks for four time periods are given in Table 2. Over time the number of edges has increased steadily, which means that more keywords are appearing together in the more recent articles. This suggests that obesity research has been diversifying over time. Furthermore, it is evident that the values of the next two network measures (i.e., average weighted degree and network density) in Table 2 show a continual increase over time. This indicates that more keywords or topics have appeared together in the obesity research and their frequencies have increased over time. The other two measures, modularity and average clustering co-efficient, have slightly decreased over time. It suggests that more keywords from different groups have been appearing together over time, indicating an increasing trend of inter-disciplinary research conducted within the obesity domain.

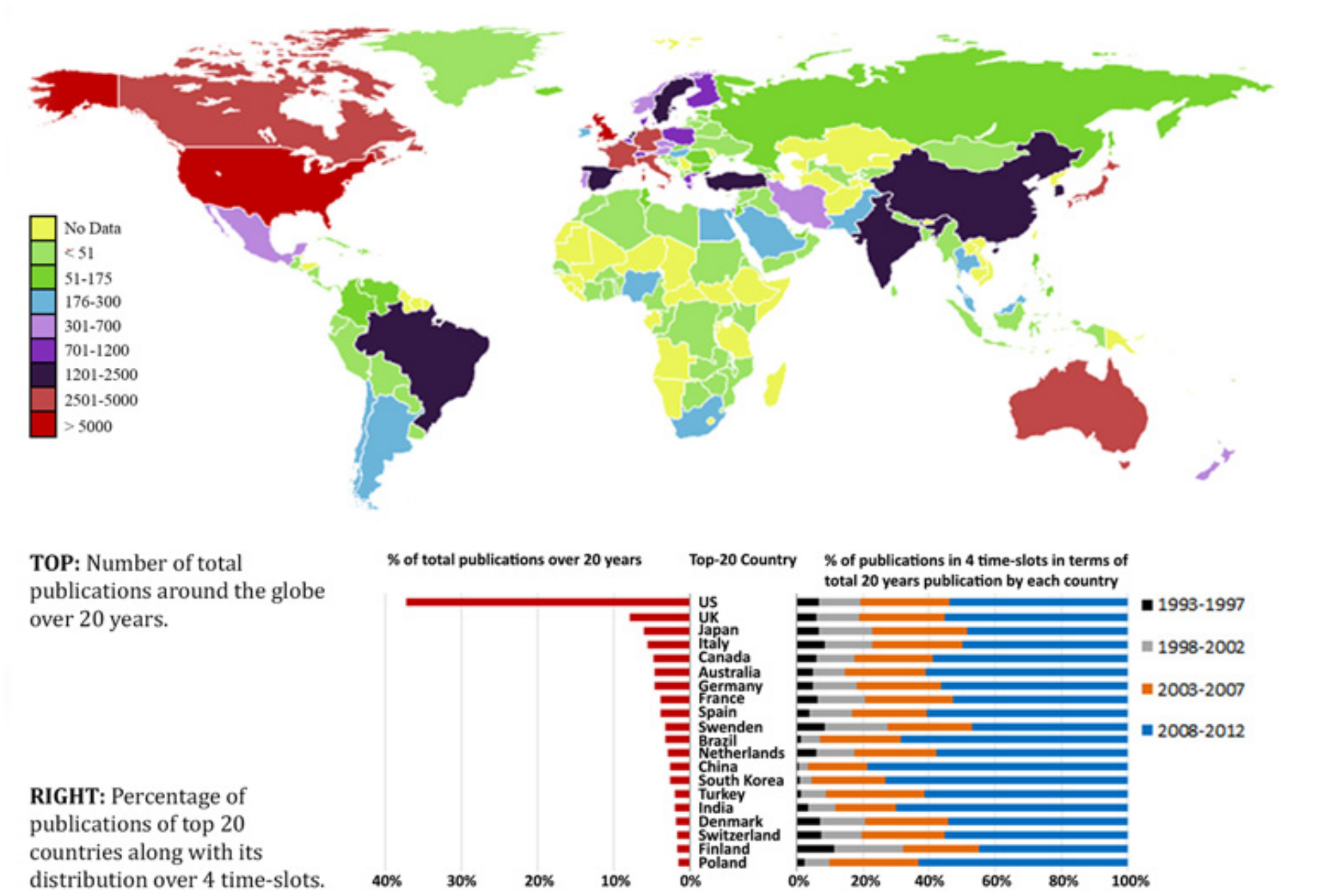

**Fig 3. Global distribution of published paper in obesity research by countries.**

## Trend in obesity knowledge structure

**Emerging keywords in obesity research.** Using Eq (1), we computed the growth of each keyword. The top-20 keywords with the highest growth are shown in Table 3. Using Eq (2), we then computed z-score for the growth of each keyword in order to identify whether or not this growth is statistically significant. All of these 20 keywords (except the last two) showed statistically significant growth. The top-20 keywords with the greatest decrease are shown in Table 4.

**Table 2. Network statistics of the keyword co-occurrence networks.**

| Network Measures | 1993–1997 | 1998–2002 | 2003–2007 | 2008–2012 | Total in 20 years |
|---|---|---|---|---|---|
| **Edges** | 7656 | 15542 | 24922 | 38345 | 58532 |
| **Average weighted degree** | 25.052 | 59.520 | 113.944 | 205.934 | 404.460 |
| **Network density** | 0.015 | 0.031 | 0.050 | 0.077 | 0.117 |
| **Modularity** | 0.325 | 0.320 | 0.323 | 0.276 | 0.289 |
| **Average clustering co-efficient** | 0.525 | 0.491 | 0.446 | 0.455 | 0.477 |

**Table 3. Top-20 keywords that showed significant growth in appearance in 20 years.**

| ID | Keyword | 1993–1997 (%) | 1998–2002 (%) | 2003–2007 (%) | 2008–2012 (%) | Relative growth (%) | Significance(z-score) |
|---|---|---|---|---|---|---|---|
| 1 | Adiponectin | 0.00 | 0.20 | 1.71 | 1.89 | 777.50 | 0.00 |
| 2 | Ghrelin | 0.00 | 0.10 | 0.89 | 0.64 | 723.47 | 0.00 |
| 3 | Asthma | 0.05 | 0.36 | 0.60 | 0.70 | 714.90 | 0.00 |
| 4 | Metabolic syndrome | 0.27 | 1.16 | 5.35 | 5.96 | 701.42 | 0.00 |
| 5 | Orlistat | 0.07 | 0.43 | 0.51 | 0.23 | 446.33 | 0.00 |
| 6 | Inflammation | 0.00 | 0.37 | 1.66 | 2.63 | 406.42 | 0.00 |
| 7 | Polymorphism | 0.17 | 0.92 | 0.77 | 0.55 | 387.69 | 0.00 |
| 8 | Nutritional status | 0.07 | 0.30 | 0.25 | 0.41 | 354.78 | 0.00 |
| 9 | TNF-alpha | 0.20 | 0.96 | 0.68 | 0.50 | 334.27 | 0.00 |
| 10 | Visfatin | 0.00 | 0.00 | 0.06 | 0.25 | 330.42 | 0.00 |
| 11 | Sleeve gastrectomy | 0.00 | 0.00 | 0.21 | 0.86 | 317.13 | 0.00 |
| 12 | Adults | 0.10 | 0.38 | 0.42 | 0.48 | 312.81 | 0.00 |
| 13 | Oxidative stress | 0.00 | 0.19 | 0.69 | 0.97 | 311.13 | 0.00 |
| 14 | Bariatric surgery | 0.69 | 2.39 | 4.35 | 3.45 | 308.58 | 0.00 |
| 15 | Primary care | 0.05 | 0.13 | 0.28 | 0.35 | 303.95 | 0.00 |
| 16 | Waist circumference | 0.25 | 0.79 | 1.21 | 1.56 | 303.20 | 0.00 |
| 17 | Adipokines | 0.32 | 0.22 | 0.78 | 1.40 | 301.92 | 0.00 |
| 18 | Pediatrics | 0.07 | 0.15 | 0.34 | 0.53 | 286.62 | 0.00 |
| 19 | Review | 0.00 | 0.00 | 0.06 | 0.22 | 277.81 | 0.01 |
| 20 | Schizophrenia | 0.00 | 0.15 | 0.56 | 0.46 | 250.79 | 0.01 |

**Table 4. Top-20 keywords that showed significant decrease in appearance in 20 years.**

| ID | Keyword | 1993–1997 (%) | 1998–2002 (%) | 2003–2007 (%) | 2008–2012 (%) | Relative growth (%) | Significance(z-score) |
|---|---|---|---|---|---|---|---|
| 1 | Norepinephrine | 0.81 | 0.28 | 0.06 | 0.00 | -243.00 | 0.02 |
| 2 | C-Peptide | 0.54 | 0.21 | 0.06 | 0.00 | -231.16 | 0.03 |
| 3 | Catecholamines | 0.71 | 0.38 | 0.10 | 0.00 | -219.31 | 0.04 |
| 4 | mRNA | 0.42 | 0.16 | 0.07 | 0.00 | -218.74 | 0.04 |
| 5 | Uncoupling protein | 0.64 | 0.38 | 0.10 | 0.00 | -214.61 | 0.05 |
| 6 | Glucose Clamp Technique | 0.32 | 0.12 | 0.06 | 0.00 | -214.24 | 0.05 |
| 7 | Plasma lipids | 0.27 | 0.12 | 0.06 | 0.00 | -207.64 | 0.05 |
| 8 | Weight cycling | 0.32 | 0.14 | 0.07 | 0.00 | -202.85 | 0.06 |
| 9 | Respiratory quotient | 0.42 | 0.32 | 0.06 | 0.00 | -202.82 | 0.06 |
| 10 | Apolipoprotein A-I | 0.10 | 0.00 | 0.07 | 0.00 | -200.00 | 0.07 |
| 11 | Bulimia | 0.12 | 0.00 | 0.07 | 0.00 | -200.00 | 0.07 |
| 12 | Hernia | 0.10 | 0.00 | 0.06 | 0.00 | -200.00 | 0.07 |
| 13 | Nutrient intake | 0.22 | 0.00 | 0.06 | 0.00 | -200.00 | 0.07 |
| 14 | Corticosterone | 0.29 | 0.19 | 0.08 | 0.00 | -193.79 | 0.08 |
| 15 | Cholelithiasis | 0.22 | 0.13 | 0.06 | 0.00 | -192.85 | 0.08 |
| 16 | Paraventricular nucleus | 0.29 | 0.16 | 0.09 | 0.00 | -191.94 | 0.08 |
| 17 | Prolactin | 0.25 | 0.16 | 0.07 | 0.00 | -191.52 | 0.08 |
| 18 | Left ventricular mass | 0.29 | 0.12 | 0.08 | 0.00 | -191.43 | 0.08 |
| 19 | Cardiac output | 0.20 | 0.09 | 0.06 | 0.00 | -190.98 | 0.08 |
| 20 | Fluoxetine | 0.32 | 0.14 | 0.09 | 0.00 | -190.49 | 0.08 |

**Table 5. Growth of frequently used keywords (second column of table 1).**

| ID | Keyword | 1993–1997 (%) | 1998–2002 (%) | 2003–2007 (%) | 2008–2012 (%) | Relative growth (%) | Significance(z-score) |
|---|---|---|---|---|---|---|---|
| 1 | Body mass index | 4.79 | 8.56 | 9.88 | 9.44 | 89.64 | 0.33 |
| 2 | Insulin resistance | 6.19 | 5.75 | 6.40 | 5.11 | -15.95 | 0.95 |
| 3 | Diabetes | 4.40 | 5.43 | 5.52 | 5.34 | 21.77 | 0.76 |
| 4 | Metabolic syndrome | 0.27 | 1.16 | 5.35 | 5.96 | 701.42 | 0.00 |
| 5 | Children | 2.31 | 3.74 | 4.10 | 5.02 | 93.85 | 0.31 |
| 6 | Hypertension | 5.14 | 5.05 | 3.93 | 3.23 | -41.63 | 0.76 |
| 7 | Type 2 diabetes | 4.25 | 4.13 | 3.82 | 3.54 | -17.72 | 0.94 |
| 8 | Morbid obesity | 4.01 | 4.80 | 6.16 | 2.23 | -15.75 | 0.96 |
| 9 | Weight loss | 5.19 | 4.83 | 3.97 | 3.10 | -46.58 | 0.73 |
| 10 | Leptin | 2.56 | 7.11 | 3.87 | 2.42 | 95.14 | 0.31 |
| 11 | Bariatric surgery | 0.69 | 2.39 | 4.35 | 3.45 | 308.58 | 0.00 |
| 12 | Adolescents | 1.30 | 1.97 | 3.22 | 3.85 | 134.29 | 0.16 |
| 13 | Risk factors | 3.56 | 3.57 | 3.38 | 3.02 | -15.93 | 0.95 |
| 14 | Physical activity | 1.45 | 2.13 | 2.90 | 3.10 | 90.01 | 0.33 |
| 15 | Adipose tissue | 2.85 | 3.02 | 2.32 | 2.43 | -12.42 | 0.98 |
| 16 | Insulin | 6.98 | 4.52 | 2.20 | 1.39 | -123.22 | 0.27 |
| 17 | Body composition | 4.15 | 3.35 | 2.50 | 1.72 | -75.90 | 0.52 |
| 18 | Epidemiology | 1.97 | 1.97 | 2.19 | 2.19 | 11.33 | 0.83 |
| 19 | Exercise | 2.51 | 2.46 | 1.93 | 2.00 | -19.89 | 0.93 |
| 20 | Inflammation | 0.00 | 0.37 | 1.66 | 2.63 | 406.42 | 0.00 |

None of these 20 keywords showed a statistically significant decrease (at $p<0.01$ level) as estimated by their z-score statistics. The growth of the top-20 keywords that appeared most frequently in the complete dataset is presented in Table 5. Interestingly, some of these keywords (e.g., *insulin resistance*, *hypertension* and *type 2 diabetes*) showed negative growth. Only three of them (i.e., *metabolic syndrome*, *bariatric surgery* and *inflammation*) showed statistically significant and positive growth as revealed by their z-score statistics. Apart from that, most of top-20 keywords have shown a steady growth and relative positions of these top-ranked keywords did not change much over time indicating constant importance of these topics over time. These keywords (e.g., *diabetes*, *children* and *hypertension*) loosely represent the broader category of different obesity domains but not a particular or specific topic. They are also from different domains and occupy considerable frequency in the overall spectrum, indicating the diversified nature of obesity research.

**Keyword categorization.** Using the MeSH database, the most frequent 650 keywords were categorized into 40 major categories. These categorized keywords comprised about 50% of total keyword appearances for our research dataset. The relative increments of each category and subcategory are given in Fig 4.

The category *"Diagnostic techniques and procedures"* includes keywords, such as, body mass index, blood pressure, weight and laparoscopy, and had the highest frequency (15% of overall frequency). This is followed by *Endocrine System Diseases* (including *diabetes*, *polycystic ovary syndrome*, *gestational cancer* etc.) and *Therapeutics* (including *bariatric surgery*, *gastric bypass*, *vertical banded gastroplasty* etc.). Less frequently used categories cover areas such as various body compounds (e.g., *lipids* and *hormones*), diseases, ethnicity, health science etc. The result shows that there is an expected strong relation between obesity and endocrine system diseases (e.g., diabetes), with a substantial amount of research having been done in this field. Next to *endocrine system diseases—metabolic diseases* and *heart diseases* have also occupied significant research attention. However, there has been relatively less work undertaken on *nutrition disorders*, *age group* and *demography* related topic in conjunction with obesity. Further details of this categorization are presented in S2 Table.

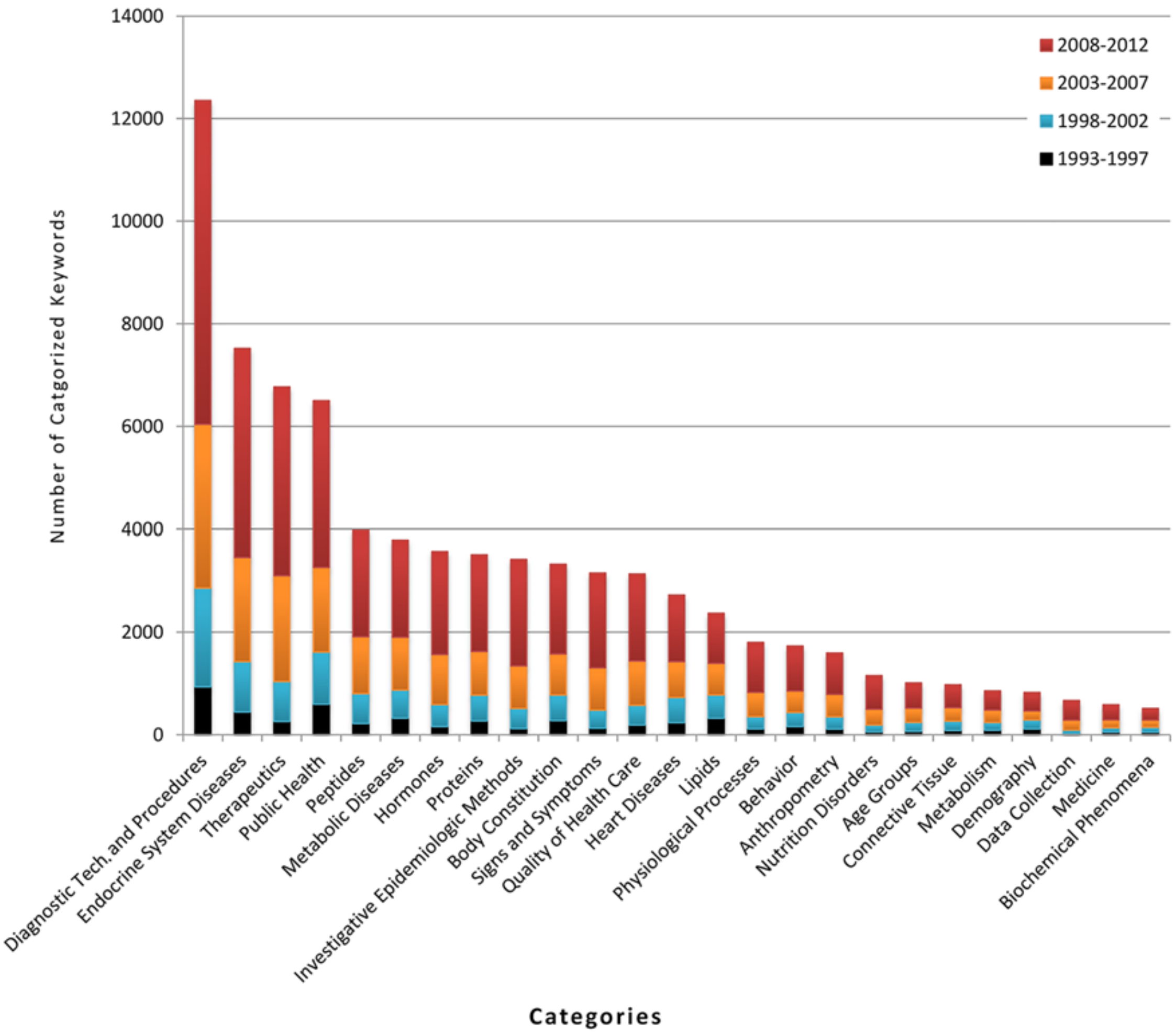

**Fig 4. Frequency distribution of categorized and sub-categorized keywords over 4 time-slots and in overall 20 years.**

**Co-occurrence and growth of keyword-pair.** The top-20 keyword pairs that showed the highest growth are presented in Table 6. All of these keyword pairs showed statistically significant growth as revealed by their corresponding z-score statistics. The growth of the most frequently appearing 20 keyword pairs in the complete dataset is illustrated in Table 7. According to their z-score statistics, there were 5 keyword pairs (i.e., *morbid obesity* and *bariatric surgery*, *body mass index* and *children*, *bariatric surgery* and *gastric bypass*, *body mass index* and *waist circumference*, and *metabolic syndrome* and *hypertension*) that showed statistically significant growth at $p < 0.01$ level.

## Knowledge map of obesity research

Fig 5 illustrates a two-dimensional knowledge map with major keywords as nodes for the complete time period (i.e., 1993–2012). It shows the frequency of keywords and their relative co-occurrence with colors of different intensity to represent it as a heat map. Intense (red) color indicates the frequency of keywords. In the center of each intense color, there is a keyword which is labelled in larger font size. This indicates that these keywords appeared more frequently and co-occurred with a higher number of other keywords in the literature. Keywords that have co-occurred more frequently are placed closer in the map. Similarly, we created knowledge maps with keywords as nodes for each of the four time slots (see S1, S2, S3 and S4 figs). Table 8 shows the keyword list that appeared with higher font size and red font color in these knowledge maps.

**Table 6. Top-20 keyword pairs that show highest growth over time.**

| ID | Keyword-1 | Keyword-2 | 1993–1997 (%) | 1998–2002 (%) | 2003–2007 (%) | 2008–2012 (%) | Relative growth (%) | Significance (z-score) |
|---|---|---|---|---|---|---|---|---|
| 1 | Bariatric surgery | Laparoscopy | 0.02 | 0.53 | 0.65 | 0.16 | 20.02 | 0.00 |
| 2 | Bariatric surgery | Gastric banding | 0.02 | 0.34 | 0.48 | 0.15 | 12.06 | 0.00 |
| 3 | Childhood obesity | Nutrition | 0.00 | 0.00 | 0.01 | 0.07 | 10.96 | 0.00 |
| 4 | Gastric banding | Sleeve gastrectomy | 0.00 | 0.00 | 0.01 | 0.06 | 10.00 | 0.00 |
| 5 | Body mass index | Adipokines | 0.00 | 0.01 | 0.01 | 0.05 | 8.56 | 0.00 |
| 6 | Metabolic syndrome | Cohort study | 0.00 | 0.00 | 0.01 | 0.05 | 7.61 | 0.00 |
| 7 | Leptin | Body fat | 0.02 | 0.22 | 0.03 | 0.02 | 7.50 | 0.00 |
| 8 | Type 2 diabetes | Gastric bypass | 0.00 | 0.00 | 0.01 | 0.10 | 7.37 | 0.00 |
| 9 | Leptin | TNF-alpha | 0.02 | 0.21 | 0.09 | 0.07 | 7.19 | 0.00 |
| 10 | Body mass index | Waist circumference | 0.05 | 0.38 | 0.62 | 0.80 | 7.18 | 0.00 |
| 11 | Complications | Sleeve gastrectomy | 0.00 | 0.00 | 0.01 | 0.05 | 7.13 | 0.00 |
| 12 | Body mass index | Children | 0.10 | 0.80 | 0.77 | 0.71 | 7.04 | 0.00 |
| 13 | Body mass index | Leptin | 0.05 | 0.41 | 0.20 | 0.13 | 6.81 | 0.00 |
| 14 | Children | Prevalence | 0.02 | 0.16 | 0.05 | 0.10 | 6.74 | 0.00 |
| 15 | Body mass index | Cohort study | 0.02 | 0.20 | 0.26 | 0.17 | 6.67 | 0.00 |
| 16 | Diet | Pregnancy | 0.00 | 0.01 | 0.01 | 0.04 | 6.65 | 0.00 |
| 17 | Leptin | Body composition | 0.02 | 0.20 | 0.06 | 0.04 | 6.61 | 0.00 |
| 18 | Adolescents | Waist circumference | 0.00 | 0.01 | 0.02 | 0.12 | 5.70 | 0.00 |
| 19 | Leptin | Glucose | 0.02 | 0.16 | 0.03 | 0.02 | 5.31 | 0.00 |
| 20 | Type 2 diabetes | Roux-en-Y gastric bypass | 0.00 | 0.00 | 0.01 | 0.04 | 5.22 | 0.00 |

**Gene-word analysis.** We found 20 gene-words in our research dataset. Cross-reference features and related keyword frequencies of these gene-words are reported in Table 9. Keywords that are derived from Obesity (e.g., *morbid obesity*, *diet-induced obesity*, *childhood obesity* etc.) had the highest average frequency (39.51). Gene-word "*Insulin*" also had a high number of derived keywords that includes *insulin resistance*, *insulin sensitivity*, *insulin secretion* etc. and these derived keywords have the third highest (14.34) average frequency after diabetes (21.99). Gene-words that work as prefix (e.g., hyper and hypo as in *hypertension*, *hypoxia*) or suffix (e.g., ~tomy as in *gastrectomy*) also had a significantly high number of derived keywords. Also, this analysis suggests that though obesity research is a huge multifaceted field, it is still evolving from a few core or "buzz" words. Around 35% of all keyword occurrences result from derived keywords that are formed by only 20 gene-words of Table 9.

## Discussion

### Research framework for exploring knowledge structure

Understanding trends and emerging topics of a research domain is indispensable for prospective and current researchers, funding agencies, policy makers and other stakeholders. However, given the tremendous amount of available bibliographic information from online sites and services, it can be extremely challenging and cognitively difficult at the same time. The primary contribution of our proposed framework, as illustrated in Fig 1, is that it can organize the scholarly information and reveal the underline longitudinal trend in a simple and intuitive way to a wide range of audiences. It gives user guidelines on downloading and organizing bibliographic data from online repositories and services. In doing so the proposed framework addresses issues like data inconsistencies, ambiguity, duplication and segmentation that are commonly noticed in different bibliographic dataset. The filtering methods introduced in the framework can potentially be applied on different online datasets regardless of the context. Our proposed framework also contributes to the present literature by proposing a set of hierarchical analysis methods that reveals generic to more specific trends of the research field. The first phase of analysis can find the research magnitude and growth which can potentially reveal the underlying factors driving the research workforce. These factors may dictate how the research problems are addressed by policy makers of different countries, funding and workforce distribution, and evolution of the research problem. The second phase of our analysis can be a powerful tool to observe micro level changes of different factors affecting the evolution of research. For instance, a particular research topic might pose a great potential to research communities and attract more workforce and sponsors (causing significant positive growth), or that topic might be exhausted since alternatives were found (causing significant negative growth). In this case, over time statistics of keywords having highest or lowest growth as well as growth of frequent keywords can show how importance have been given or changed on this research topic. The categorization scheme introduced in this phase can be a guide to group keywords into logical categories. Applying the same set of analysis on these grouped keywords, macro level trends of research sub-domains can be explored. The third phase of analysis can help to understand how different research topics are interrelated and evolve (e.g., co-occurrence test and knowledge map). It can also help to comprehend the knowledge structure of the domain through hybridization and cross-reference of few stem keywords (e.g., Gene-word analysis).

**Table 7. Top-20 keyword pairs that have been found most time in 20 years and their growth.**

| ID | Keyword-1 | Keyword-2 | 1993–1997 (%) | 1998–2002 (%) | 2003–2007 (%) | 2008–2012 (%) | Relative growth (%) | Significance (z-score) |
|---|---|---|---|---|---|---|---|---|
| 1 | Morbid obesity | Bariatric surgery | 0.52 | 2.13 | 3.22 | 0.94 | 4.75 | 0.00 |
| 2 | Morbid obesity | Gastric bypass | 0.81 | 0.95 | 1.62 | 0.38 | 0.04 | 0.90 |
| 3 | Insulin resistance | Metabolic syndrome | 0.07 | 0.22 | 1.12 | 0.83 | 1.93 | 0.06 |
| 4 | Children | Adolescents | 0.22 | 0.66 | 0.58 | 0.90 | 2.49 | 0.02 |
| 5 | Body mass index | Children | 0.10 | 0.80 | 0.77 | 0.71 | 7.04 | 0.00 |
| 6 | Bariatric surgery | Gastric bypass | 0.10 | 0.60 | 1.16 | 0.55 | 4.95 | 0.00 |
| 7 | Diabetes | Hypertension | 0.76 | 0.84 | 0.77 | 0.60 | -0.18 | 0.94 |
| 8 | Body mass index | Waist circumference | 0.05 | 0.38 | 0.62 | 0.80 | 7.18 | 0.00 |
| 9 | Insulin resistance | Type 2 diabetes | 0.79 | 0.68 | 0.84 | 0.43 | -0.51 | 0.70 |
| 10 | Body mass index | Adolescents | 0.15 | 0.52 | 0.58 | 0.55 | 2.52 | 0.01 |
| 11 | Morbid obesity | Laparoscopy | 0.42 | 1.21 | 0.95 | 0.17 | 0.76 | 0.42 |
| 12 | Diabetes | Metabolic syndrome | 0.00 | 0.10 | 0.63 | 0.60 | 0.01 | 0.92 |
| 13 | Insulin resistance | Diabetes | 0.29 | 0.68 | 0.57 | 0.43 | 0.99 | 0.31 |
| 14 | Leptin | Insulin | 0.42 | 1.24 | 0.50 | 0.23 | 0.52 | 0.57 |
| 15 | Metabolic syndrome | Hypertension | 0.05 | 0.23 | 0.59 | 0.45 | 3.58 | 0.00 |
| 16 | Risk factors | Epidemiology | 0.42 | 0.38 | 0.44 | 0.42 | -0.11 | 0.99 |
| 17 | Weight loss | Bariatric surgery | 0.07 | 0.27 | 0.58 | 0.37 | 2.35 | 0.02 |
| 18 | Hypertension | Blood pressure | 0.44 | 0.48 | 0.37 | 0.38 | 0.05 | 0.89 |
| 19 | Morbid obesity | Weight loss | 0.44 | 0.49 | 0.75 | 0.18 | -0.53 | 0.69 |
| 20 | Morbid obesity | Gastric banding | 0.32 | 0.79 | 0.69 | 0.13 | 0.59 | 0.52 |

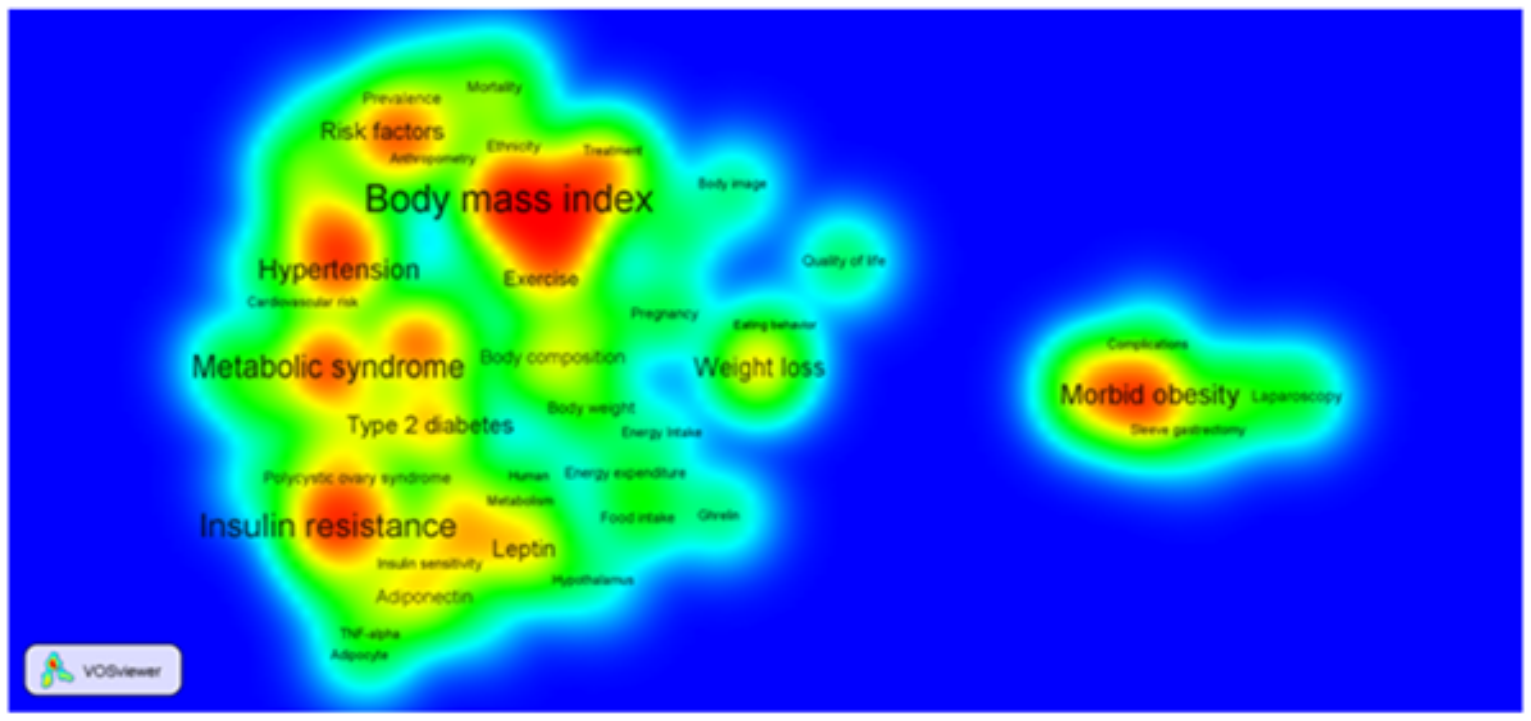

**Fig 5. Visualization of two-dimensional knowledge map of keywords for the complete dataset (i.e., 1993–2012).**

**Table 8. Keywords that appeared with higher font size and red font color in different knowledge maps.**

| Time period | Keyword list |
|---|---|
| 1993–1997 | Blood pressure; Hypertension; Insulin; and Morbid obesity |
| 1198–2002 | Body composition; Body mass index; Diabetes; Insulin resistance; Leptin; Morbid obesity; and Weigh loss |
| 2003–2007 | Body mass index; Hypertension; Insulin resistance; Metabolic syndrome; and Morbid obesity |
| 2008–2012 | Bariatric surgery; Body mass index; Insulin resistance; Metabolic syndrome; Risk factors; and Waist circumference |
| 1993–2012 (20 years) | Body mass index; Hypertension; Insulin resistance; Leptin; Metabolic syndrome; Morbid obesity; Risk factors; and Weight loss |

**Table 9. Gene-words and their appearance statistics over time.**

| Gene-word | Number of Derived Keywords | Frequency (1993–1997) | Frequency (1998–2002) | Frequency (2003–2007) | Frequency (2008–2012) | Frequency (20 years) | Average frequency of each keyword |
|---|---|---|---|---|---|---|---|
| Obesity | 860 | 2164 | 4463 | 8541 | 18810 | 33978 | 39.51 |
| Diabetes | 324 | 412 | 971 | 1905 | 3837 | 7125 | 21.99 |
| Insulin | 567 | 912 | 1393 | 2297 | 3530 | 8132 | 14.34 |
| Adipose | 486 | 369 | 745 | 1627 | 3852 | 6593 | 13.57 |
| Metabolism/ Metabolic | 424 | 170 | 398 | 1486 | 3501 | 5555 | 13.1 |
| Child | 542 | 186 | 533 | 1374 | 3565 | 5658 | 10.44 |
| Surgery | 370 | 120 | 376 | 1082 | 2039 | 3617 | 9.78 |
| Cardio | 363 | 175 | 408 | 901 | 2000 | 3484 | 9.6 |
| Weight | 818 | 534 | 1139 | 2107 | 3939 | 7719 | 9.44 |
| Hyper | 644 | 562 | 1048 | 1637 | 2736 | 5983 | 9.29 |
| Risk | 567 | 252 | 623 | 1252 | 2474 | 4601 | 8.11 |
| Lipid | 372 | 174 | 433 | 771 | 1413 | 2791 | 7.5 |
| Blood | 290 | 208 | 383 | 566 | 1006 | 2163 | 7.47 |
| Disease | 725 | 230 | 580 | 1340 | 3153 | 5303 | 7.31 |
| Diet | 746 | 388 | 737 | 1316 | 2551 | 4992 | 6.69 |
| Food | 537 | 120 | 255 | 545 | 1395 | 2315 | 4.31 |
| Hypo | 417 | 132 | 321 | 471 | 768 | 1692 | 4.06 |
| ~tomy | 343 | 64 | 116 | 304 | 909 | 1393 | 4.06 |

To summarize, we proposed a framework that can effectively organize online bibliographic data and perform a set of analysis to understand the knowledge structure of research evolution. Note that the analysis methods that are presented here are not unique. They are well established in statistical and computer science domain and have been implemented effectively on various contexts. However, combining them into a framework in a hierarchical fashion to summarize knowledge structure from online bibliographic data can arguably be considered as a unique and novel approach.

## Observation from obesity research using the proposed research domain

The analysis of knowledge structure based on the last 20 years of scholarly articles on obesity can give a big picture of what has been done as well as suggest future research directions. The finding from the basic analysis shows increasing number of articles over years that indicate growing involvement of research community around the globe and funding to tackle the obesity epidemic. In fact, publication numbers doubled in each of the five years periods (see Fig 2). However, this increment is not uniform across countries. The United States, for example, has outnumbered other countries in the number of research publications and has shown steady growth over time. The recent rise in obesity research publications in countries such as China, India and Turkey reflects the rising prevalence of obesity in many low and middle income countries which are undergoing economic and nutrition transition [30].

The keyword set showing the highest growth (see Table 3) normally points to specific obesity research fields that have gained research interest in recent years. For example, two keywords with highest growth—*adioponectin* and *ghrelin*—were discovered in 1995 [40] and 1996 [41] respectively. It then took several years to gain importance in the scientific community, resulting in a high growth. Also, some keywords (e.g., *Asthma* and *Adults*) had high growth or importance because of changing socio-economic or epidemiological factors [31]. Finally, the keyword set showing lowest (i.e., negative) growth (see Table 4) indicates that the research on that area is exhausted, or is not producing effective result, or some more effective keyword have replaced those topics. For example, this study found that keywords *orlistat* and *sibutramine* (anti-obesity drug) both had a similar frequency (appeared 200 and 201 times respectively in the 20 years of publication data) but the first one had positive growth (446%) while the latter had negative growth (-59%). This is analogous with the fact that *sibutramine* is associated with increased cardiovascular events (and strokes) and has been withdrawn from most countries' markets while *orlistat* is considered a safer drug. A similar effect can be seen in the options for *bariatric surgery*.

From the categorization analysis, it is evident that most research has been undertaken in categories that contain diseases or disorder related keywords. This is followed by almost the same proportion of research on explicit indicators (e.g., behavioral, socio-economic and external risk factors of obesity) and complex internal processes (e.g., biological, hormonal and molecular) that influence energy balance. Also, categorization shows that research targeting specific age groups (having keywords *children*, *adolescents*, *elderly*, *adults*, *youth* etc.) the number is relatively low compared to other categories such as diagnostic procedures, metabolic diseases, body constitution etc. On the other hand keyword "*Children*" from age group category is the fifth highest occurring keyword, emphasizing the rising prevalence of childhood obesity and associated research.

The co-occurrence network of keywords shows a high clustering co-efficient (see Table 2). In fact, the co-efficient score of 0.477 is higher than many real world networks. For example, the co-authorship network of a specific scientific domain, airline connection network and neural network have clustering co-efficient score of 0.44, 0.34 and 0.15 respectively [42]. This indicates that the topics of obesity research are highly correlated and studied together. Also, if we look at highly co-occurred keywords (see Fig 4 and Table 7) we can see several keywords (e.g., *insulin resistance*, *metabolic syndrome*, *type 2 diabetes* and *hypertension*) occur together more frequently in all possible pairs, between them forming a tightly knitted cluster in the co-occurrence map. Furthermore, co-occurrence test and the visualization show at which point different obesity sub-domains merge. For example, the keyword *Insulin resistance* belongs to the diabetes related field and *polycystic ovary syndrome* is one type of endocrine disorder of female. These two keywords from somewhat different domains co-occur significantly (see Table 7 and Fig 4) in obesity research indicating that obesity is the common field where those two domains have merged.

Finally, we wish to discuss the quality and accuracy of results that we found by applying the proposed framework to the obesity domain. The dataset used in this research were collected from Scopus which is the most reliable and provides the largest database of peer-reviewed materials. We only considered the journal articles and did not include 2013 publications as this list was still incomplete at the time of data collection. Some articles were not published in English and we excluded those from our dataset as including those would incur complexity in translating foreign keywords into English. These excluded articles are very few compared to our dataset size, and their omission is unlikely to have a major impact on our results. With regard to the quality of keywords, we chose all the keywords that were provided by the actual authors of those articles and did not use any controlled vocabulary. On average, each article had around five author-provided keywords to describe itself. Furthermore, as noted in the Methods section, we filtered keywords extensively by merging synonymous keywords and correcting misspelled entities. In addition, the categorization scheme followed in this research was developed by the US National Library of Medicine which is comprehensive and especially designed to index articles of medical science. Therefore, we can argue that, the dataset used in our framework is extensive, and relatively complete and accurate within an acceptable boundary. It is, however, possible that research articles in some of the social sciences may have been missed, due to the use of Scopus as a bibliographic database rather than, say, Google Scholar [43].

## Conclusion

In this paper, we have presented a novel framework for exploring knowledge structure of any given research domain. Although we applied this framework only to the obesity research domain, it can be utilized to explore the knowledge structure of any other research domains. As multi-disciplinary domains such as obesity involve researchers, practitioners, policymakers and others from diverse backgrounds, there is a greater importance for a framework that can present research trends in an intuitive and user-friendly way. Our framework for exploring a knowledge structure exactly targets to this issue. Such a knowledge structure allows stakeholders to make various decisions which may otherwise be difficult using traditional methods. The analyses presented in this paper are not exhaustive. So far, we have given most of our endeavor to effective organization of the huge amount of unfiltered data and have performed several analyses such as gene-word, co-occurrence tests on keywords and visual mapping. However, several important entities in our knowledge system still remain to be analyzed. For example, funding information in our dataset could be linked to other indicators (e.g., citation count) and keywords to find out how funding affects the overall quality of research and which sub-domains of obesity have higher proportions of different funding sources. Relating funding details to affiliation can also suggest which institutions receiving major research grants. This may help aspiring researchers to select prominent institutions to pursue studies or to choose research topic wisely. For policy makers, it can also help them to understand the effect of funding and make effective future plans. Another approach can be to apply advanced social network and graph-theoretic approaches to the dataset, including performing cluster test, centrality tests on keywords and affiliation data. This may reveal otherwise hidden group structures of the keywords and show bridging topics between different research domains. Finally, author's affiliation (institution, department, state, country etc.) can be analyzed and put on a geographical map to give readers a visual cue as to which institutions or areas are at the forefront of particular areas of obesity research and how they collaborate.

## Supporting Information

**S1 Fig. Visualization of two-dimensional knowledge map of keywords for 1993–1997.**
(PNG)

**S2 Fig. Visualization of two-dimensional knowledge map of keywords for 1998–2002.**
(PNG)

**S3 Fig. Visualization of two-dimensional knowledge map of keywords for 2003–2007.**
(PNG)

**S4 Fig. Visualization of two-dimensional knowledge map of keywords for 2008–2012.**
(PNG)

**S1 Table. Top-50 frequently appeared keywords in the research dataset.** Keywords with grey backgrounds appeared in all four time periods (could be in different numerical order).
(DOCX)

**S2 Table. Major keywords under each category.**
(DOCX)

## Acknowledgments

The authors like to express their sincere gratitude to Jeremy Cullis, faculty liaison librarian of University of Sydney medical science library for helping out in keyword categorization. Also the authors would like to thank Liang Wang who worked as a summer scholar with us and helped to download and filter the dataset.

## Author Contributions

Conceived and designed the experiments: SU AK. Performed the experiments: AK. Analyzed the data: AK SU LB. Contributed reagents/materials/analysis tools: AK SU. Wrote the paper: AK SU LB.